# Calibration of the Total Solar Irradiance Data Record

Leif Svalgaard (leif@leif.org)
Stanford University
2018-12-21

**Abstract**

Solar surface magnetic field seems to be able to explain variations in Total Solar Irradiance on timescales from hours to decades. Using magnetograms from spacecraft (MDI and HMI) and ground-based observatories (MWO and WSO) I build a composite dataset of the Total Line-of-Sight Unsigned Magnetic Flux over the solar disk stretching back to 1976, validated by excellent correlations with the solar microwave flux (F10.7) and the Sunspot Group Number. Direct measurements of TSI by space borne sensors have been carried out since late 1978. The early instruments were plagued by scattered light entering the aperture, but this construction flaw can be corrected for. At the AGU 2018 meeting, a new TSI composite has been proposed based on a novel mathematical method vetted by representatives from all current and most past TSI instruments. Although an 'official' release of the dataset has not been offered yet, a preliminary version is available. Anticipating that any last-minute changes might be minor, I compare this new version with the magnetic flux composite. It is clear that we have two TSI populations: values before 1993 that are seriously too low and values from 1993 onwards. I elect to normalize the magnetic flux (the driver of variations of TSI) to a New TSI using the regression equation for the recent population with the smallest uncertainty. With this normalization, there is now total agreement between the variation of the magnetic flux and of the New TSI as well as with the F10.7 and Group Number proxies. We now have two choices: (1) the Sun underwent a dramatic change in how its magnetic field drives variation of TSI or (2) the New Consensus TSI reconstruction does not work and the new dataset is premature and not useful neither for solar nor for climate research. Following David Hume, we should always believe whatever would be *the lesser miracle*, which in our case would be choice (2).

**Introduction**

It is well-established (e.g. Shapiro et al. [2017]; Yeo et al. [2014, 2017]) that solar surface magnetic field and (on time scales of hours) granulation can together precisely explain (what they call) 'solar noise' in Total Solar Irradiance (TSI) on timescales from minutes to decades, i.e. ranging over almost seven orders of magnitude in period. This accounts for all timescales that have so far been resolved or covered by irradiance measurements. They demonstrate that no other sources of variability are required to explain the data. The remarkable agreement between models and measurements demonstrate that our understanding of the mechanisms responsible for TSI variability is fundamentally correct, at least on timescales up to several decades.

Direct measurements of TSI by space borne sensors have been carried out since late 1978. The early instruments (before the launch of SORCE TIM, Rottman [2005]) were plagued by scattered light entering the aperture (e.g. [Kopp & Lean 2011]). Due to overlapping records from different spacecraft, this construction flaw could be corrected for. At the recent AGU 2018 meeting, a new TSI composite has been proposed (Kopp et al., AGU 2018, SH32B-08; based on a method by DuDok de Wit et al. [2017]) vetted by the TSI community (representatives from all current and most past TSI instruments) as that recommended for use by solar and climate researchers. It comes with a suitable number of buzz-words, like: using time-varying weightings of all the available instrument data rather than selecting only one instrument at any given time and relying on daisy-chaining (which propagates errors in time), using a data-driven noise model whose uncertainties are estimated in a systematic way, providing a less-biased statistical approach for selecting instrument weightings, and bridging data gaps smoothly via a multi-scale decomposition approach, providing time-dependent uncertainties. Although an 'official' release of the dataset has not been offered yet, a preliminary version is at Greg Kopp's website https://spot.colorado.edu/~koppg/TSI/Thierry_TSI_composite.txt , complete with the following warning: "this composite should not be used for publication". Anticipating that any last-minute changes might be minor, I'll compare this new version with the magnetic flux observed since 1976. This comparison shall be updated when the official release appears.

**The Magnetic Flux**

NASA's magnetograph, HMI [Schou 2012], on its 'flagship' SDO spacecraft which was launched 11 February, 2010 with the motto "all the sun, all the time" has yielded virtually complete coverage of the magnetic field in the solar photosphere. I calculate for each full disk magnetogram the total line-of-sight (LOS) unsigned (i.e. the magnitude) magnetic flux over the solar disk. Although the true flux is not accurately known (e.g. Riley et al. [2013]) it is possible to reduce all measurements to an arbitrary common scale (chosen to be HMI) due to significant overlap between instruments, with the solar microwave flux (F10.7, Svalgaard [2016]) serving as independent plausible verification. The first step is to construct a spacecraft-derived composite between HMI and its predecessor MDI (on SoHO using calibration level 1.8.2; http://soi.stanford.edu/sssc/progs/mdi/calib.html ; [Scherrer et al. 1995]). Figure 1 shows the monthly means of the MDI-HMI composite:

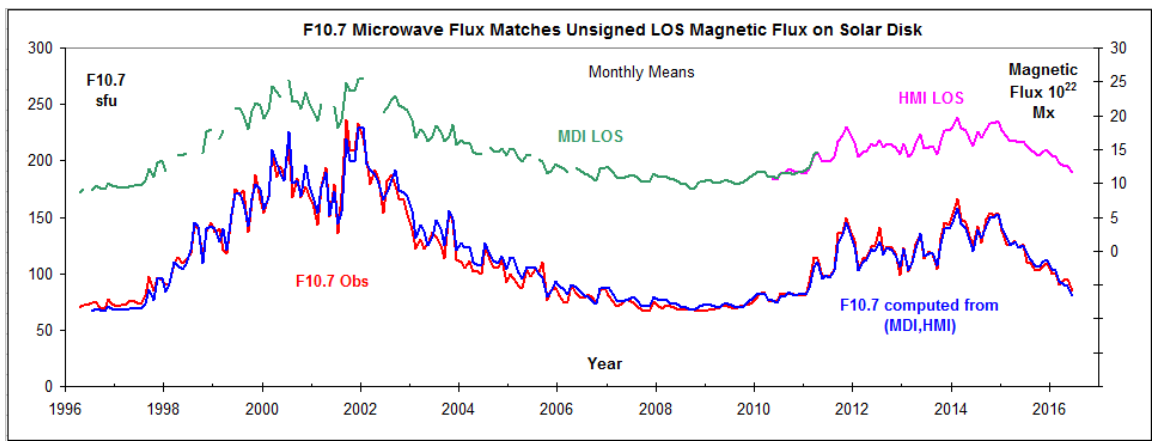

**Figure 1:** Upper green curve: MDI* LOS flux scaled to match the HMI LOS flux based on their overlap. Upper pink curve: HMI. Lower red curve: observed F10.7 flux reduced to 1 A.U. Blue curve: F10.7* computed from the magnetic flux of the composite record. The magnetic flux includes a noise component that shows up as an offset when comparing with other data. By requiring a match during the overlap the offset becomes irrelevant.

The MDI and HMI data overlap during 2010, May through 2011, April and we scale MDI to HMI ($MDI^* = 0.7429\ MDI - 2.847$) and form the composite data set as the average of HMI and the scaled MDI. The monthly averages F10.7 can be computed from the flux table at http://www.spaceweather.gc.ca/ reduced to 1 A.U. (without the URSI correction factor). There is a good, almost linear relationship between the monthly values of F10.7 and of our composite magnetic flux, $M$. For comparison purposes, we remove a slight non-linearity by fitting the relationship to a 4th degree polynomial, calculating $F10.7^* = 0.0071136\ M^4 - 0.488305\ M^3 + 12.315337\ M^2 - 125.4736\ M + 510.854$ sfu ($R^2 = 0.97$) as the microwave flux corresponding to a given magnetic flux. Part of $M$ is due to noise as we do not apply any threshold on the magnetic field. Our analysis is not affected, however, if the background is stable. The agreement between F10.7* and the magnetic flux extends to daily values ( http://hmi.stanford.edu/hminuggets/?p=1510 ) as well. This suggests that the microwave flux is controlled directly by the magnetic field without a complicated intermediate physical process and that the microwave flux is a good proxy for the total LOS unsigned magnetic flux over the solar disk. I thus consider it established that we have a good representation of the magnetic flux from the two spacecraft instruments back to 1996 as plausibly validated by the well-observed microwave flux.

Ground-based magnetographs reach further back in time, but some are plagued by 'upgrades' to the instruments that destroy the homogeneity and long-term stability of the resulting datasets. The Mount Wilson Observatory (MWO) underwent major upgrades in 1982 (instrument) and 1986 (processing). I compute the total unsigned magnetic flux over the solar disk from MWO magnetograms from 1986 through 2013 [online at ftp://howard.astro.ucla.edu/pub/obs/magfits_daily_average/ ; and John Boyden, personal communication] and scale it to the MDI*-HMI composite, Figure 2:

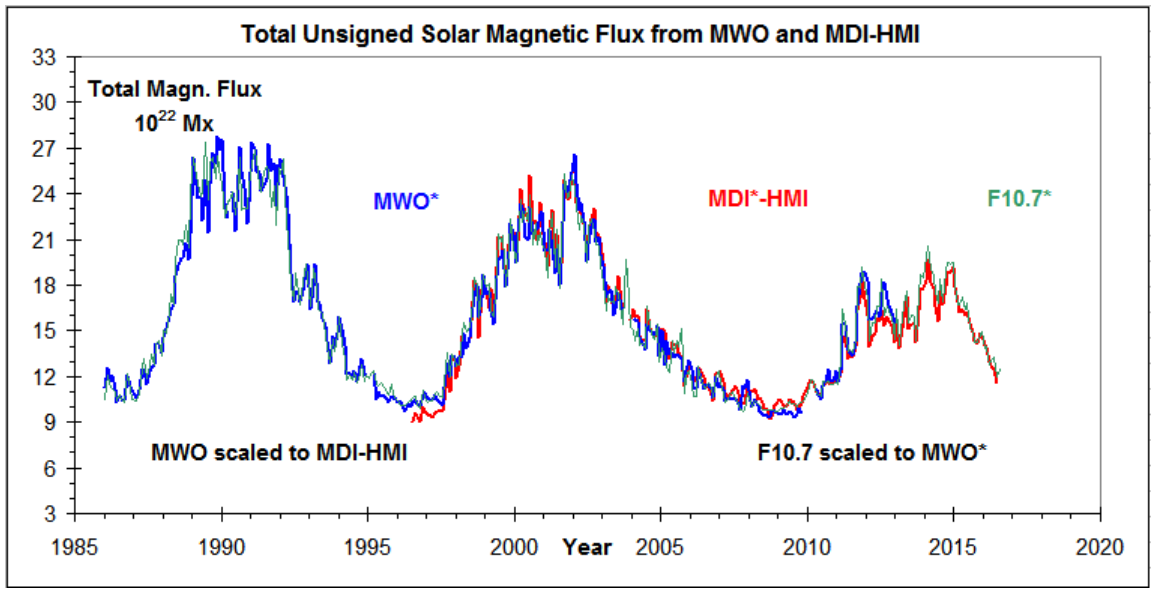

**Figure 2:** Blue curve: MWO LOS total flux scaled to match the MDI-HMI composite (red curve) validated by the F10.7 flux scaled to MWO (green curve). The F10.7 comparison is important as MDI is calibrated to MWO onboard the spacecraft. All curves have 1-month resolution.

The MWO magnetic flux fits the spacecraft-derived magnetic flux well as validated by the fit of the F10.7 flux. For each magnetogram a Magnetic Plage Strength Index (MPSI) value is also calculated ( http://obs.astro.ucla.edu/150_data.html#plots ) by summing the absolute values of the magnetic field strengths for all pixels where the absolute value of the magnetic field strength is between 10 and 100 gauss. This number is then divided by the total of number of pixels (regardless of field strength) in the magnetogram: ftp://howard.astro.ucla.edu/pub/obs/mpsi_data/index.dat yielding an alternate measure of the solar magnetic flux, Figure 3.

The Wilcox Solar Observatory (WSO) has since May 1976 made daily low-resolution maps of the Sun's magnetic field ( http://wso.stanford.edu/ ). The observatory is still operating in its original configuration ensuring a homogeneous dataset with stable calibration. As for MWO, I calculate the total unsigned magnetic flux over the disk for the ~14,000 magnetograms and scale it to the MDI-HMI composite, Figure 3:

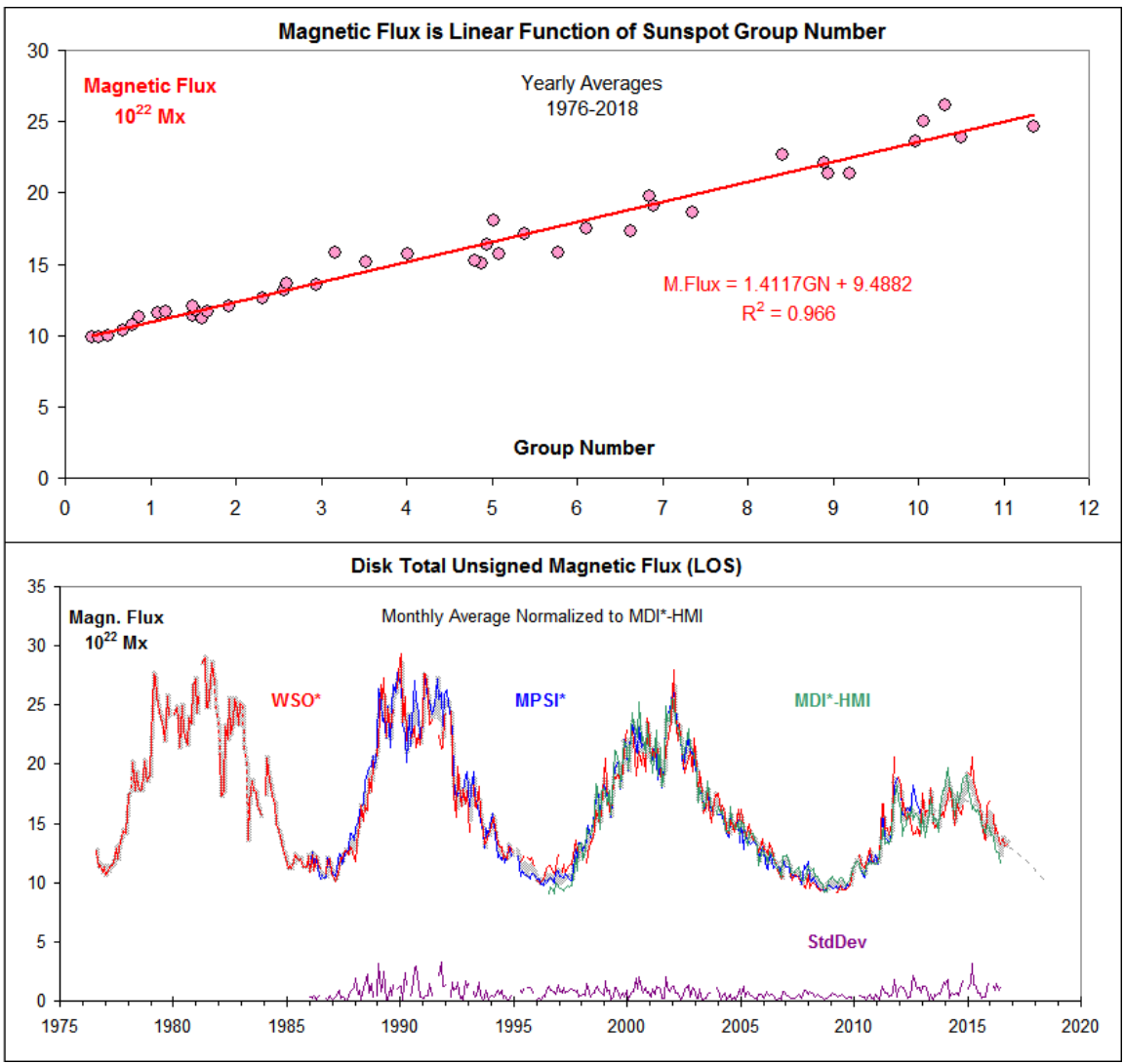

**Figure 3:** Lower panel: The WSO LOS total unsigned magnetic flux (red curve) scaled to match the MDI-HMI composite (green curve) with 1-month resolution. The MPSI from MWO (blue curve) nicely bridges the WSO and MDI-HMI composite and lends credibility to the validity of the overall variation (average of all curves; grey curve). Upper panel: The yearly average solar LOS total unsigned magnetic flux as a function of the Sunspot Group Number.

The yearly average magnetic flux closely ($R^2$ = 0.966) follows linearly the run of the Sunspot Group Number [Svalgaard & Schatten 2016]. So, again, we must conclude that we have a very good idea of how the solar magnetic flux has varied during the last four sunspot cycles. At the least, the variability of the magnetic record matches those of the proxies F10.7 and Sunspot Group Numbers. The yearly values can be found online at https://leif.org/research/magnflux.txt .

**The New Consensus TSI Record**

Verifying the assumption that solar irradiance variability is driven by its surface magnetism has been hampered by the fact that models of solar irradiance variability based on solar surface magnetism have in the past been calibrated to observed variability. Making use of realistic 3D-MHD simulations of the solar atmosphere and state-of-the-art solar magnetograms from SDO, Yeo et al. [2017] present a model of TSI that does not require any such calibration. The modeled irradiance *variability* is entirely independent of the observational record and is claimed to replicate 95% of the observed variability (SORCE/TIM) between April 2010 and July 2016, leaving little scope for alternative drivers of solar irradiance variability time scales examined (days to years).

I have pointed out for years that the SORCE/TIM TSI record appeared to be drifting in the sense that the instrument reported a steady upwards divergence from the proxies and other TSI instruments [e.g. https://leif.org/research/EUV-F107-and-TSI-CDR-HAO.pdf Slides 49 ff]. The New Consensus Series appears to have fixed that, Figure 4:

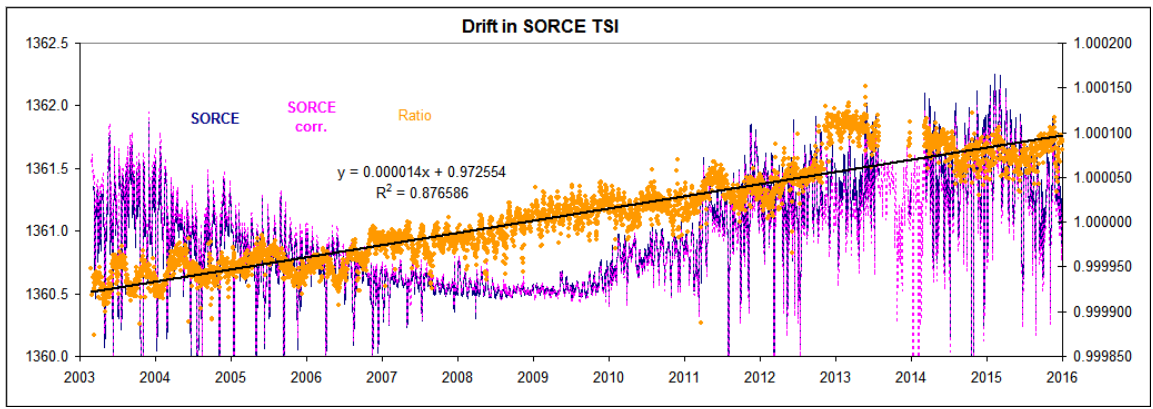

**Figure 4:** Daily values of the original (that is what is reported as version 17 on the LASP website at http://lasp.colorado.edu/data/sorce/tsi_data/daily/sorce_tsi_L3_c24h_latest.txt (blue curve) and of the New Consensus series (pink curve). The ratio between the two series is plotted as orange points and largely confirms the upward drift (with some gruesomeness in 2013 before the instrument [temporarily] failed). The drift between April 2010 and July 2016 is too small to significantly impact the Yeo et al. [2017] analysis.

With that out of the way, it is time to compare the New and Improved Consensus TSI record with the reliable magnetic flux record, Figure 5:

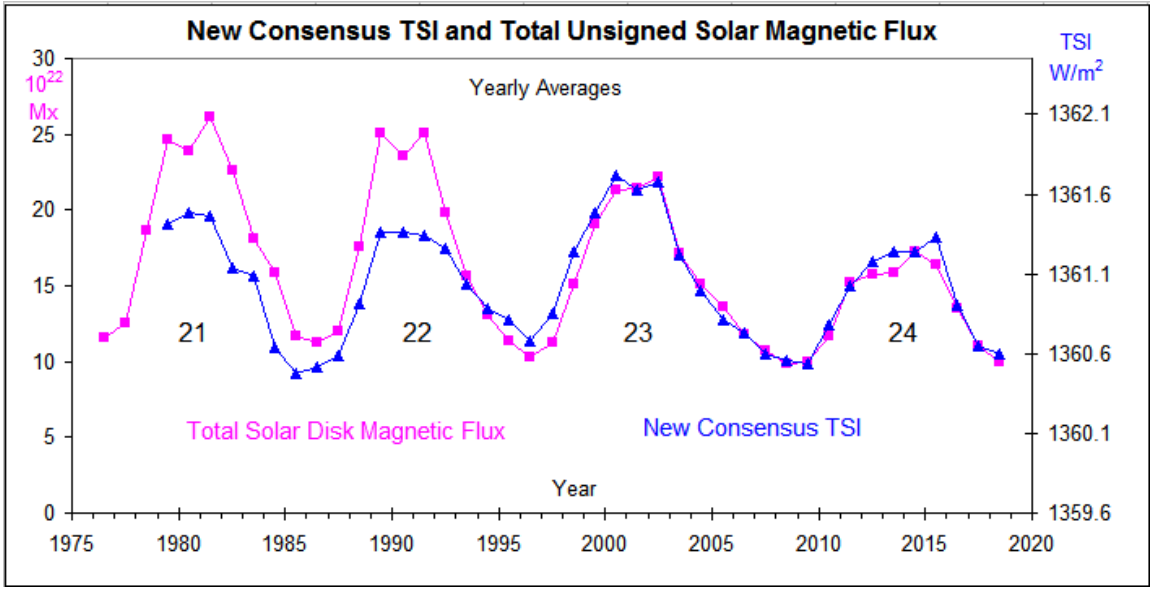

**Figure 5:** Yearly averages of the New Consensus TSI series (blue curve with triangles) for solar cycles 21 through 24 compared with the total LOS unsigned magnetic flux over the solar disk scaled to match TSI since 1993 (pink curve with squares).

If we plot the New TSI against the solar magnetic flux (Figure 6) it is clear that we have two populations: values before 1993 (red triangles) and values from 1993 on (blue dots). As the uncertainty is smallest for the more recent population I elect to normalize the magnetic flux (the driver of variations of TSI) to a New TSI using the regression equation for the recent population, Figure 7.

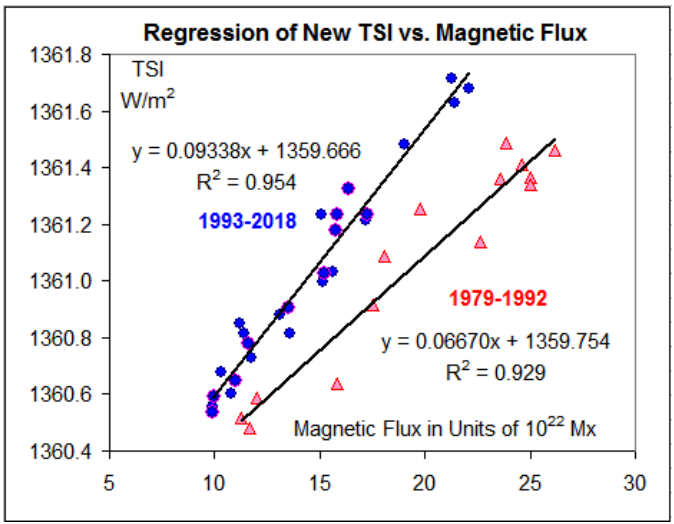

**Figure 6:** Yearly values of New TSI vs. total LOS solar magnetic unsigned flux in units of $10^{22}$ Mx. Values before 1993 are shown as red triangles, while values from 1993 onwards are shown as blue dots (for solar cycle 24 with a red outline). The scatter is (as expected) larger for the early data points.

With this normalization, there is now total agreement between the variation of the magnetic flux and of the normalized New TSI as we would expect from the Yeo et al. [2017] analysis. In particular, the sizes of solar cycles 21 and 22 as measured by TSI are now in line with all other solar measures for those cycles. One could argue that with a stated absolute uncertainty of 0.5 W/m² for the early New TSI data, the divergence was still (barely) within the error-band, but this seems to take the form of special pleading.

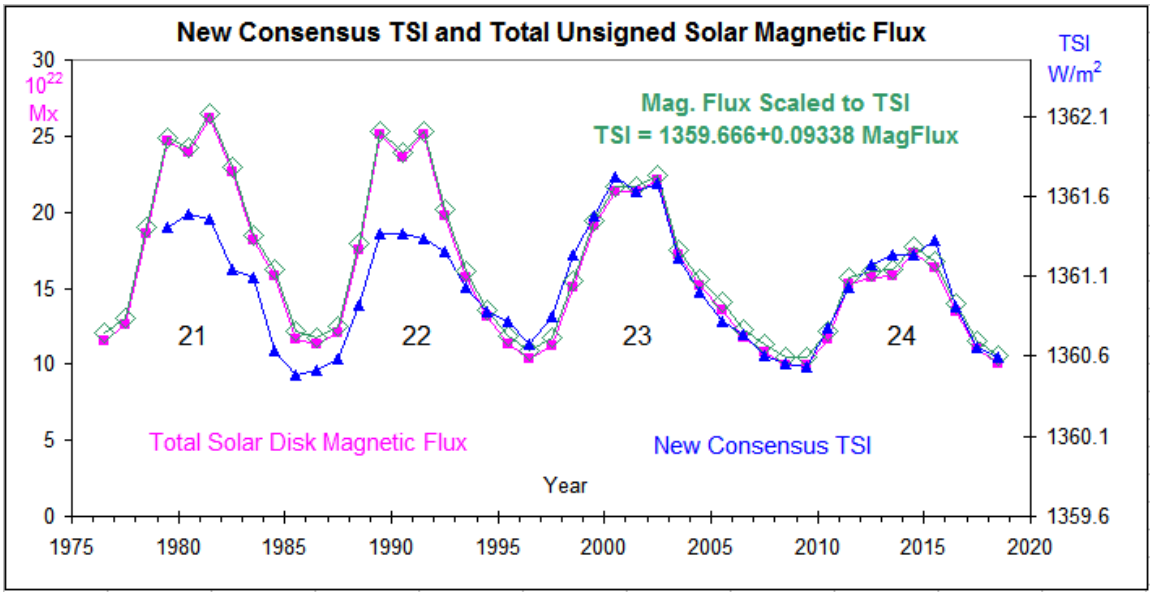

**Figure 7:** Yearly averages of the New Consensus TSI series (blue curve with triangles) for solar cycles 21 through 24 compared to the total LOS unsigned magnetic flux over the solar disk scaled to match TSI since 1993 (pink curve with squares). The TSI that would result from normalizing the magnetic flux to TSI using the equation of Figure 6 for the most recent two cycles with the smallest uncertainty is shown as the green curve with open diamonds.

**The Lesser Miracle**

We now have two choices: (1) the Sun underwent a dramatic change beginning around the start of the year 1993 (perhaps ~1993.092 = 1993/02/03) in how its magnetic field drives variation of TSI or (2) the New Consensus TSI reconstructed using the DuDok de Wit et al. [2017] methodology does not work as intended and that therefore the New Consensus TSI dataset is premature and not useful in climate research. David Hume, in Section X of his *Enquiry Concerning Human Understanding* [1748], argued that a rational person should never believe that a miracle (he is using the word 'miracle' in the everyday sense, meaning something that is merely out of the ordinary) had actually taken place unless it would be a greater miracle that the person reporting the miracle was simply mistaken. We should always believe whatever would be *the lesser miracle*, which in our case would be choice (2), suggestive of a return to the drawing board.

**Acknowledgement**

I thank Stanford University for continuing support, and Phil Scherrer for helpful comments.